\documentclass[letter]{aa} % for the letters 
\usepackage{color}

\usepackage{graphicx}
\usepackage{txfonts}
\usepackage{epstopdf}
\begin{document}

   \title{Tip of the red giant branch distance to the nearby dwarf galaxy [TT2009]\,25 in the NGC\,891 group}
   \author{
          Oliver M\"uller\inst{1}
          \and
                    Rodrigo Ibata\inst{1}
                    \and
          Marina Rejkuba\inst{2}
          \and  
          Lorenzo Posti\inst{1}
          }

 \institute{Observatoire Astronomique de Strasbourg  (ObAS),
Universite de Strasbourg - CNRS, UMR 7550 Strasbourg, France\\
 \email{oliver.muller@astro.unistra.fr}
\and
 European Southern Observatory, Karl-Schwarzschild Strasse 2, 85748, Garching, Germany
}
   \date{Received tba; accepted tba}

% \abstract{}{}{}{}{} 
% 5 {} token are mandatory
  \abstract{Dwarf galaxies are key objects for small-scale cosmological tests like the abundance problems or the planes-of-satellites problem. It is therefore a crucial task to get accurate information for as many nearby dwarf galaxies as possible. Using extremely deep, ground-based $V$ and $i$-band Subaru Suprime Cam photometry with a completeness of $i=27$\,mag, we measure the tip of the red giant branch distance for the dwarf galaxy [TT2009]\,25. This dwarf resides in the field around the {Milky Way-analog} NGC\,891. By using a Bayesian approach, we measure a distance of $10.28^{+1.17}_{-1.73}$\,Mpc, which is consistent with the distance of NGC\,891, thus we confirm it as a member of NGC\,891. The dwarf galaxy follows the scaling relations defined by the Local Group dwarfs. {We do not find an extended stellar halo around [TT2009]\,25.} In the small field of view of 100\,kpc covered by the survey{, only one bright dwarf galaxy and the giant stream are apparent. This is comparable to the Milky Way, where one bright dwarfs reside in the same volume, as well as the Sagittarius stream} -- excluding satellites which are farther away but would be projected in the line-of-sight. It is thus imperative to survey for additional dwarf galaxies in a larger area around NGC\,891 to test the abundance of dwarf galaxies and compare it to the number of satellites around the Milky Way. } 

   \maketitle
%
%________________________________________________________________

\section{Introduction}
In the $\Lambda$CDM universe structure is build up through hierarchical merging in a bottom-up scenario \citep[e.g.\ the review by ][]{2012AnP...524..507F}. The smallest building blocks are the dwarf galaxies, which are also the most abundant and most dark matter dominated {galaxies} in the universe. In recent years, using the dwarf galaxies, several tests 
%for
{of} cosmological models have been proposed and conducted, scrutinizing the abundance, the distribution, the motions, and the internal dynamics of the dwarfs \citep[e.g., ][]{1999ApJ...524L..19M,2010A&A...523A..32K,2018NatAs...2..925K,2011MNRAS.415L..40B,2014Natur.511..563I,2016ApJ...817...75L,2018MNRAS.481..860R,2018Sci...359..534M,2018Natur.555..629V,2019A&A...623A..36M,2019ApJ...870...50J,2019MNRAS.487.2441H}. For some of these tests the models have successfully found a solution (e.g.\ \citealt{2007ApJ...670..313S,2016MNRAS.457.1931S,kroupa2018does,2019MNRAS.484.1401R}), others are still {actively discussed in the literature} (e.g.\ \citealt{2015ApJ...815...19P,2015MNRAS.452.1052L,2018MPLA...3330004P,2019arXiv190307285M,2019arXiv190703761M}). To further improve and extend these tests, it is crucial to measure these properties for as many dwarf galaxies in as many different environment as possible \citep{2016A&A...588A..89J,2017ApJ...847....4G}.

In the nearby universe, one of the most effective ways to estimate the distance of a galaxy is via the tip of the red giant branch (TRGB, \citealt{daCostaArmandroff1990,Lee1993}).
It exploits the maximal brightness a {low-mass} star can reach along its evolutionary path in %its
{the} red giant branch {(RGB)} phase {when observed in appropriate optical or near-infrared filters}. 
%This cut-off 
{The observed sharp increase in the ($I$-band) luminosity function, or cut-off in the colour-magnitude diagram (CMD), which is theoretically understood \citep{2017A&A...606A..33S} as well as empirically calibrated \citep{2008MmSAI..79..440B},} can be successfully used as a standard candle, yielding distance 
{measurements accurate to about}
%accuracies of 
5 to 10 percent \citep{2015ApJ...802L..25T,MuellerTRGB2018,2018ApJ...868...96C,2019arXiv190511416A,2019arXiv190603230B}.
{In $I$-band the observed TRGB depends only weakly on} 
%mostly independent of 
age and metallicity, and {is typically not affected by} the presence of variable stars {provided sufficiently well sampled stellar population}.

In this work, we present a TRGB distance measurement for the dwarf galaxy [TT2009]\,25, discovered by \citet{2009MNRAS.398..722T} in a wide-field MegaCam based survey of the region around NGC\,1023, and more specifically close to the Milky Way analog NGC\,891. The Milky Way analog NGC\,891 has been targeted  to resolve its stellar {halo and search for streams through detection of}
individual {RGB} stars \citep{2010ApJ...714L..12M} with ground-based Subaru data, accompanied with data from the Hubble space telescope \citep{2009MNRAS.396.1231R}. While the previous work focused on {the stellar halo, which has revealed interaction history with a smaller companion that left} streams and loops around NGC\,891, in the following we will measure the distance to the dwarf galaxy [TT2009]\,25, which resides within the deep Subaru field.

\section{Data and photometry}

\begin{figure*}[ht]
\includegraphics[width=18cm]{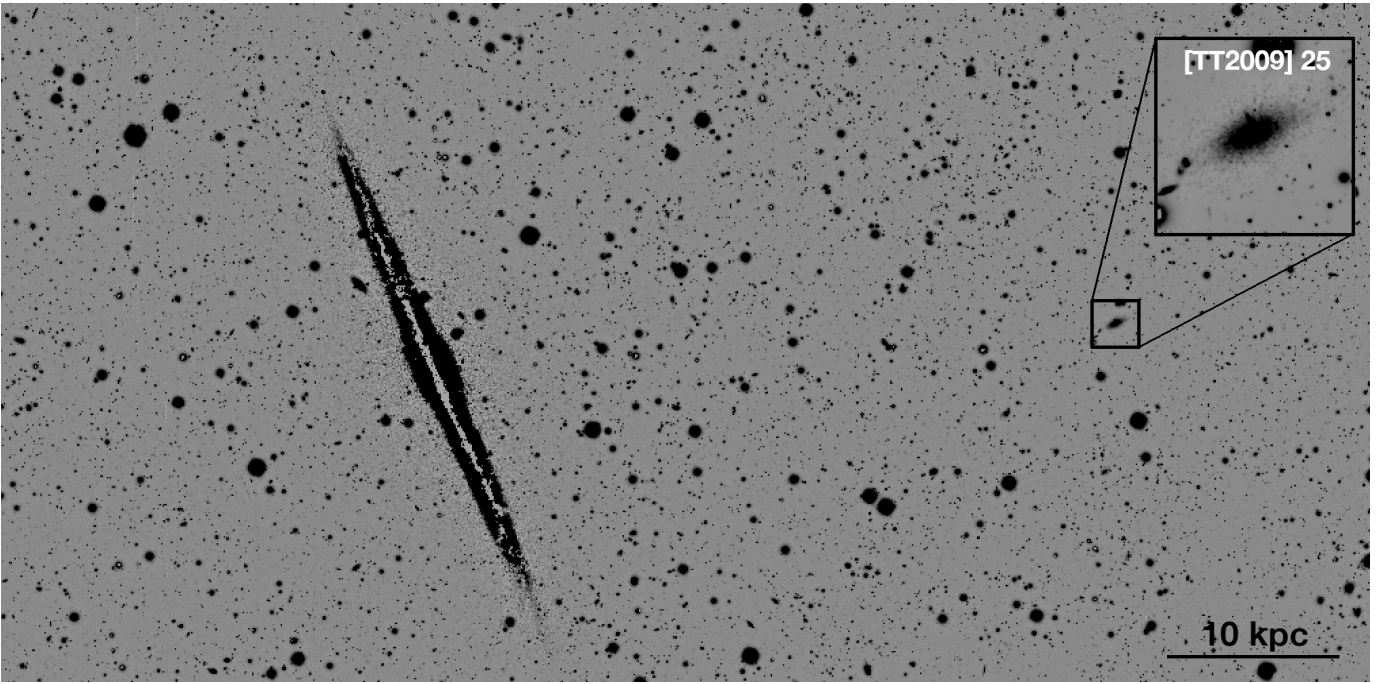}
\caption{The field around NGC\,891 from our deep, stacked $i$-band Subaru Suprime Cam image. The box highlighting the dwarf galaxy [TT2009]\,25 has a side of 1\,arcmin. The scale indicates 10\,kpc at 10\,Mpc.}
\label{field}
\end{figure*}

\begin{figure}[ht]
\centering
\includegraphics[width=9cm]{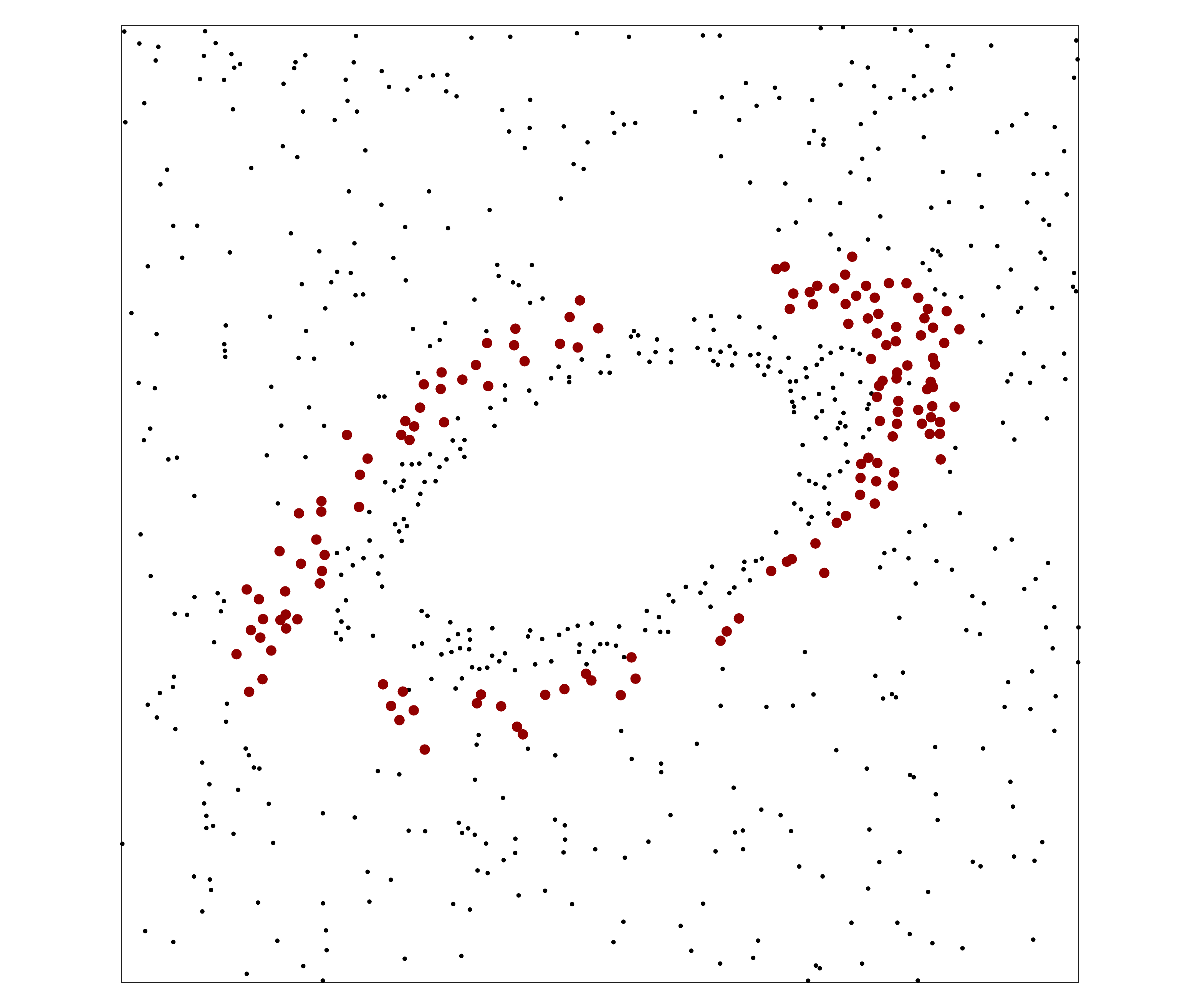}
\caption{Stars map around [TT2009]\,25. Indicated are objects fainter than $i=25.5$\,mag. The large red dots correspond to the stars selected for the color magnitude diagram. The inner part of the galaxy is obfuscated due to the crowding of stars. One side is 100 arcsec.}
\label{field_TT}
\end{figure}

{In this work we use data from the study of NGC\,891 by \citet{2010ApJ...714L..12M} based on deep Subaru Suprime Cam \citep{2002PASJ...54..833M} imaging. 
The observations were conducted in the Johnson $V$-band and the Gunn $i$-band. A total of 10\,hours of good quality data in the $V$-band and 11 hours in the $i$-band were obtained. The seeing was better than 0.6\,arcsec. This strategy resolved the RGB down to 2 magnitudes below the tip at the distance of 10 Mpc \citep{2007MNRAS.381..873M}, corresponding to NGC\,891. 

The photometry was performed with DAOPHOT \citep{1987PASP...99..191S}. For the detailed description of the photometric pipeline we refer to \citet{2010ApJ...714L..12M}. 
We have calibrated the photometry with the Pan-STARRS DR2 catalog \citep{2016arXiv161205560C}. While calibrating our $i$-band was a straight forward task, we had to convert  Pan-STARRS $gr$ to our $V$ using \citet{SloanConv}.
The photometric errors in the $i$-band {range from} $i\approx0.05$\,mag at $i\approx25.8-26.0$\,mag, i.e., the bright end of the RGB, to $i\approx0.15$\,mag at  $i\approx26.8-27.0$\,mag \citep{2010ApJ...714L..12M}, i.e., approximately a magnitude below the TRGB of NGC\,891. This is sufficient to accurately measure the distance at the 5 to 10 percent level. To cull bona-fide stars from background compact galaxies we have applied quality cuts using the {PSF fitting} sharpness and $\chi^2$ parameters, and rejected objects deviating from the mean photometric errors (see e.g. Fig.\,2 of \citealt{2019arXiv190702012M}).

\section{TRGB distance measurement}

In Fig\,\ref{field} we present a cut-out of the field around NGC\,891, including the dwarf. It is clearly visible and its outskirts are well resolved in stars, while in the centre the crowding gets too high for accurate photometry.} 
In Fig\,\ref{field_TT} we show the star map of putative RGB stars. To create a color magnitude diagram (CMD), we select all stars within an elliptical annulus, avoiding the innermost stars. This was necessary as blending of individual stars was evident, i.e. their magnitudes were  $\sim$0.4\,mag higher than in the outskirt of the galaxy. We refer to \citet{2019MNRAS.486.1192T} for a recent discussion about the effect of crowding in dwarf galaxies.

We measure the TRGB magnitude {using} a Bayesian Markov Chain Monte Carlo (MCMC) approach developed by \citet{2011ApJ...740...69C,2012ApJ...758...11C} {and described in detail in \citet{2019arXiv190702012M}}. In this scheme, we approximate the  {location of RGB stars on the CMD} with a power law, and model the contamination of foreground/background objects {using} a large reference field. The reference field consists of several patches avoiding the stellar loop and the inner halo of NGC\,891, with roughly the same distance to NGC\,891 as [TT2009]\,25, with an area of $\approx$100 square arcmin.
A most likely set of $m_{TRGB}$, $a$, and $c$ are evaluated, where  $m_{TRGB}$ is the TRGB magnitude, $a$ the slope of the power law, and $c$ the contamination factor.

\begin{figure}[ht]
\includegraphics[width=8.9cm]{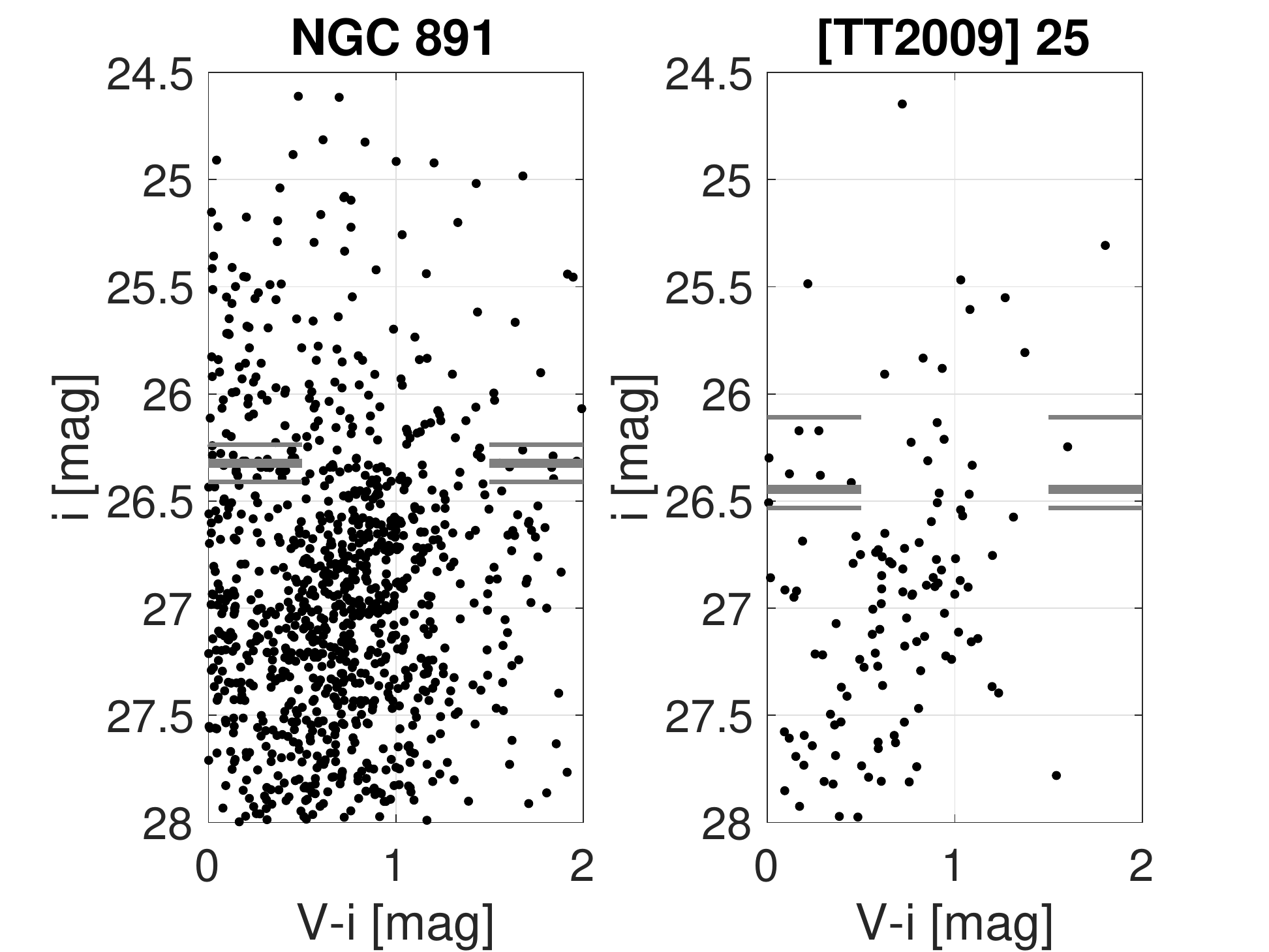}
\caption{The color magnitude diagram of NGC\,891 (left) and [TT2009]\,25 (right). The gray lines denote the best estimated TRGB magnitude and the corresponding uncertainties.}
\label{CMD}
\end{figure}

We estimate the distance to [TT2009]\,25 by measuring the difference in the distance modulus between NGC\,891 and [TT2009]\,25. This has the advantage that we can anchor our estimate to the distance of NGC\,891 coming from the Hubble space telescope, yielding more accurate distances than what is possible from ground based telescopes{, due to the superior image quality}. 
For NGC\,891 we use an elliptical annulus placed such that it only covers the outskirts of the galaxy. Using our MCMC scheme, we measure $i_{TRGB}=26.32^{+0.09}_{-0.09}$\,mag. The error bounds are given by the 68\% interval of the MCMC chain and  already include the photometric uncertainty. This gives a distance of $D=8.97^{+0.36}_{-0.36}$\,Mpc for NGC\,891 using the calibration of the TRGB magnitude of $-3.44\pm0.1$ (\citealt{2008MmSAI..79..440B}, transformed into AB magnitudes), which indeed differs slightly from the HST distance of $D=9.73^{+0.94}_{-0.86}$\,Mpc \citep{2007MNRAS.381..873M,2009MNRAS.396.1231R}.
For [TT2009]\,25 we measure the TRGB to be at $i_{TRGB}=26.45^{+0.09}_{-0.34}$\,mag. %, i.e. a distance estimate of $D=8.11^{+1.42}_{-0.42}$\,Mpc. 
In Fig\,\ref{CMD} we present the CMD for NGC\,891 and the dwarf galaxy [TT2009]\,25, as well as the estimated TRGBs. The difference in the distance modulus between NGC\,891 and [TT2009]\,25 is $0.12$\,mag, however, within the uncertainties the two values are consistent. Using a distance of $D=9.73$\,Mpc for NGC\,891 \citep{2007MNRAS.381..873M,2009MNRAS.396.1231R} this yields the final extinction corrected distance estimate of $D=10.28^{+1.17}_{-1.73}$\,Mpc for [TT2009]\,25, with a distance modulus of $(m-M)_=30.06^{+0.23}_{-0.40}$\,mag. {The uncertainties come from the joint errors of the TRGB detection in the Subaru data ($^{+0.12}_{-0.35}$\,mag) and the uncertainty in the TRGB detection in the HST data ($\pm0.2$\,mag).}

\section{Discussion}
In the following we discuss the structural properties of [TT2009]\,25 and the abundance of dwarf galaxy members around NGC\,891.
\subsection{Characterizing [TT2009]\,25}
{The distance estimate of [TT2009]\,25 is consistent with it being a %member 
{satellite} of NGC\,891.} 
This is not surprising, as the velocity $v=692.0\pm58$\,km/s of [TT2009]\,25 \citep{2015Ap.....58..309K} is also consistent with that of NGC\,891 ($v=526\pm7$\,km/s, \citealt{1987MNRAS.224..953S}). 
Nonetheless, with the distance estimation we confirm beyond doubt that [TT2009]\,25 is a dwarf galaxy %member of
{associated with}
NGC\,891.

Having the distance information, we can derive its structural parameters. For this task we use Galfit \citep{2002AJ....124..266P} -- a standard program to measure galaxy profiles. Due to the dwarf being resolved, we used MegaCam data from the CFHT archive, where the dwarf is unresolved. We downloaded the reduced $r$-band data, calibrated the image using Pan-STARRS DR2 \citep{2016arXiv161205242M}, and performed aperture photometry  with DAOPHOT to measure the zero point of the image. The dwarf galaxy was  then fit with a S\'ersic model \citep{1968adga.book.....S} using Galfit. The derived parameters are shown in Table \ref{properties}.

The structural parameters follow the relation defined by the dwarfs in the Local Group \citep{2008ApJ...684.1075M,2012AJ....144....4M} and beyond \citep[e.g. ][]{2017A&A...608A.142V,2017A&A...597A...7M,2018A&A...615A.105M}, and are reminiscent of the dwarf {spheroidal} galaxy Fornax \citep{1995MNRAS.277.1354I} in absolute magnitude, effective radius, and ellipticity. {The analogy to Fornax goes only so far as [TT2009]\,25 has emission in the FUV in GALEX \citep{2005ApJ...619L...1M}, indicating star formation. So [TT2009]\,25 is rather a transition type dwarf then a dwarf spheroidal.}
Furthermore, no {atomic hydrogen is clearly detected with deep radio interferometry} at the position of this dwarf galaxy \citep{2007AJ....134.1019O}, which is expected for bound dwarf galaxies around giants \citep{2014ApJ...795L...5S}. 

{The stellar profile of [TT2009]\,25 seems to have a rather sharp cut-off (visible in Fig\,\ref{field_TT}). This argues against an extended stellar halo around this dwarf. The existence or absence of such extended stellar halos has been discussed in the literature \citep[e.g., ][]{1996ApJ...467L..13M,2000AJ....120..801R,2003Sci...301.1508M,2004ApJ...614L.109G,2009ApJ...705..704H, 2018ApJ...861...81G}, with growing evidence that  even low mass galaxies may have experienced hierarchical growth (e.g. \citealt{2016ApJ...826L..27A}) and could thus host extended stellar envelopes. However,} {the superposition of the outer halo population of NGC\, 891 in front of [TT2009]\,25 would require deeper data to detect any such low surface brightness  envelope around the dwarf galaxy (if present).}

As a final note, [TT2009]\,25 seems not to be associated with the giant stream around NGC\,891. This is evident when studying Fig.\,1 of \citet{2010ApJ...714L..12M}, where the RGB stars in the full field are displayed {and [TT2009]\,25 is well visible as seperated overdensity of stars}. The dwarf galaxy seems to be well {away} from the stream {and $\sim 500$~kpc behind the halo of NGC\, 891}.

\begin{table}[!htb]
\caption{Properties of [TT2009]\,25.}% title of Table     % is used to refer this table in the text
\centering                          % used for centering table
\begin{tabular}{l c }        % centered columns (4 columns)
\hline\hline                 % inserts double horizontal lines
& [TT2009]\,25  \\    % table heading
\hline      \\[-2mm]                  % inserts single horizontal line
RA (J2000)& 02:21:12.2 \vspace{1mm}\\ 
DEC (J2000)& $+$42:21:50 \vspace{1mm}\\
%$i_{TRGB}$ (mag) & $25.50^{+0.06}_{-0.07}$   \\
$(m-M)_0$ (mag) &  $30.06^{+0.23}_{-0.40}$  \vspace{1mm}\\
Distance (Mpc) & $10.28^{+1.17}_{-1.73}$\vspace{1mm}\\
$A_i$ (mag)& 0.09\vspace{1mm}\\
%$[$Fe/H$]$ (dex) & $-1.79\pm0.4$ \\
$m_r$ (mag) & $16.88\pm0.06$ \vspace{1mm}\\
$M_{r}$ (mag) & $-13.18^{+0.24}_{-0.40}$  \vspace{1mm}\\
$L_{r}$ ($10^6$ M$_\odot$) & $16.0^{+7.1}_{-3.2}$  \vspace{1mm}\\
$r_{eff,r}$ (arcsec) & $13.02\pm0.04$ \vspace{1mm}\\
$r_{eff,r}$ (pc) & $649^{+76}_{-111}$ \vspace{1mm}\\
$\mu_{eff,r}$ (mag arcsec$^{-2}$)& $24.45\pm0.06$ \vspace{1mm}\\
$PA$ (north to east) & $116.9\pm0.1$\vspace{1mm}\\
$e$ $(1-b/a)$ &  $0.52\pm0.01$\vspace{1mm}\\
S\'ersic index $n$ & $1.01\pm0.01$\\
\hline
\end{tabular}
\label{properties}
\end{table}

\subsection{Abundance of dwarf galaxies around NGC\,891}
Apart from the giant stream surrounding NGC\,891, [TT2009]\,25 is the only resolved object in our field. This is 
%insofar unsurprising
{not unexpected}, as the field of view only covers $\sim$100\,kpc at the distance of NGC\,891. 
%This giant galaxy 
The spiral galaxy {NGC\,891} is thought to be a Milky Way analog in size and mass. When we take the dwarf galaxy satellites around the Milky Way within a sphere of 50\,kpc and brighter than $-10$\,mag \citep{2012AJ....144....4M,2013MNRAS.435.1928P}, we find only the Sagittarius dSph and the Large Magellanic Cloud, i.e.\ two satellites. This is %hence comparable 
{quite comparable}
to what we find for NGC\,891: {a stream that is reminiscent of Sagittarius and a moderately bright dwarf galaxy}. At fainter magnitudes though, the Milky Way hosts at least 10 dwarf galaxies \citep{2012AJ....144....4M}, with the numbers still increasing. It is thus to be expected that many low mass and low surface brightness satellites await discovery within the observed field around NGC\,891. Indeed,  \citet{2009MNRAS.395..126I} identified 6 overdensities within the much smaller HST/ACS area imaged close to the main galaxy disk, that could be either candidate satellites or satellite debris. {In a wider field of view, \citet{2015AstBu..70..379K} found no new dwarf galaxy candidate based on a 12 hour exposure with an amateur telescope, apart of [TT2009]\,25 and [TT2009]\,30. This does not mean that NGC\,891 is sparsely populated. Within 300\,kpc, the Local Volume catalog \citep{2004AJ....127.2031K,2013AJ....145..101K} yields four more supsected irregular type dwarf galaxies based on their radial velocities:  DDO\,22, DDO\,24, UGC\,1807 and UGC\,2172.}

\section{Summary and conclusion}
Using extremely deep $V$ and $i$ band images taken with the Subaru Suprime Cam we have resolved the upper part of the red giant branch of the dwarf galaxy [TT2009]\,25. We have measured the tip of the red giant branch {brightness} using a Markov Chain Monte Carlo method and have derived an accurate distance for this dwarf galaxy, which is consistent with the distance of NGC\,891. We thus confirm its membership. 

Using data from the CFHT archive, we have measured the structural parameters of [TT2009]\,25, which are compatible in absolute magnitude, effective radius, and ellipticity to other known dwarf galaxies. Most notably, it is reminiscent of the Local Group dwarf {spheroidal} galaxy Fornax {in morphology, even though the detection of FUV emission makes [TT2009]\,25 rather a transitional type dwarf. Still,} {this makes the dwarf galaxy an interesting target for follow-up studies. For example, the globular cluster distribution around Fornax yields strong constraints about the shape of its dark matter halo, favoring a core-like profile over a cusp \citep{2012MNRAS.426..601C,2019MNRAS.485.2546B,2019MNRAS.tmp.1624O}. Identifying globular clusters around [TT2009]\,25 could therefore bring new insights into the cusp/core problem \citep{2010AdAst2010E...5D}, as well as directly measuring the velocity dispersion profile, which should be possible with future facilities like the Thirty Meter Telescope.}

Due to the limited field of view, we only cover a region of 50\,kpc around NGC\,891. In this field, we find one resolved dwarf galaxy. This is comparable to what is expected from the satellite distribution around the Milky Way, where two bright/classical dwarf galaxies reside within a radius of 50\,kpc. However, this does not include dwarf galaxies which can be farther away from the host, but projected along our line-of-sight, which would increase the number of expected satellites observed around NGC\,891. It is thus imperative to search for dwarf galaxies in a larger area around NGC\,891 to test the abundance of dwarf galaxies and compare it to the Milky Way. 

\begin{acknowledgements} {We thank the referee for the constructive report, which helped to clarify and improve the manuscript.}
O.M. is grateful to the Swiss National Science Foundation for financial support. The authors thank Filippo Fraternali for discussions about the HI content in the field, {and Michele Bellazzini for discussion about the RGB tip calibration in different filter systems}. 
\end{acknowledgements}

\bibliographystyle{aa}
\bibliography{aanda}

\end{document}